\documentclass{JFM-FLM_Au}
 \usepackage{amsmath}
\usepackage{subcaption}
\usepackage{xcolor}
\usepackage{xspace}
\usepackage{orcidlink}
\usepackage{hyperref}
\usepackage{float}
\usepackage{enumitem}

\newcommand{\uad}{\boldsymbol{u}^{\dagger}}
\newcommand{\upe}{\boldsymbol{u}{'}}

\lefttitle{P. Keuchel, D. Mor\'on and M. Avila}

\title{Scaling of the minimal energy for turbulence transition in pipe flow}

\author{Patrick Keuchel\aff{1},
  Daniel Mor\'on \aff{1}\aff{2}
 \and Marc Avila \aff{1}\aff{3}
 \corresp{\email{marc.avila@zarm.uni-bremen.de}}
 }

\affiliation{\aff{1} Center of Applied Space Technology and Microgravity (ZARM), University of Bremen, Am Fallturm 2, 28359 Bremen, Germany
\aff{2} Department of Aerospace Engineering, Escuela Politécnica Superior, Universidad Carlos III de Madrid, 28911 Leganés, Spain
\aff{3} MAPEX Center for Materials and Processes, University of Bremen,
Am Biologischen Garten 2, 28359 Bremen, Germany
}

\begin{document}

\maketitle

\begin{abstract}
Predicting the transition to turbulence in pipe flow remains a fundamental problem in fluid dynamics.
We use a variational approach to compute nonlinear optimal perturbations to the laminar flow. 
Optimal perturbations are structurally similar for all Reynolds numbers, $1750 \leq Re \leq 10^4$, considered. As $Re$ increases, they increasingly localise in space with thickness scaling as $ Re^{-1/3}$.
They grow via the Orr mechanism, followed by a phase of energy transfer across oblique waves of different axial wavenumber, characterised by a rapid delocalisation. The energy gain in the Orr and oblique--wave phases is independent of the initial perturbation energy, $E_0$. Subsequently, rolls grow by nonlinear mechanisms and streaks via the lift--up effect.
The energy gain of the (inviscid) Orr phase is independent of the Reynolds number, whereas it increases linearly with $Re$ during the initial oblique--wave phase and quadratically in the roll and lift--up phase.
We find that, transition occurs if the energy of the optimal perturbation exceeds a maximum energy threshold during its temporal evolution. This energy is given by the edge state energy and found to be independent of $Re$.
As a result, the minimum perturbation energy required to cause transition in pipe flow scales as $E_0^\text{crit}\propto Re^{-3}$, which is consistent with our computation of minimal seeds.
We note that insufficiently long pipes lead to an exponent of smaller magnitude and that much larger numerical resolutions are required to accurately compute the energy gain than to compute the critical threshold.
\end{abstract}

\begin{keywords}
Authors should not enter keywords on the manuscript.
\end{keywords}

\section{Introduction}
The characteristics of laminar and turbulent flows are fundamentally different, and understanding and predicting the transition between these states has been a central problem in fluid dynamics since the pioneering work of \citet{Reynolds1883}.
Steady pipe flow is governed by the Reynolds number, $Re=RU_c/\nu$, where $R$ is the pipe radius, $U_c$ the centreline velocity of the laminar profile and $\nu$ the kinematic viscosity. 
Once triggered, turbulence in long pipes is sustained provided that $Re\gtrsim 2040$ \citep{Avila2011}. Predicting the transition to turbulence, however, is extremely difficult. This difficulty stems from the linear stability of the system up to at least $Re\gtrsim 10^7$ \citep{Meseguer2003}, whereas transition in experiments is observed in the range $1500 \lesssim Re \lesssim 10^5$ \citep{Pfenniger1961}.
\\
This discrepancy between linear stability and experimental observations is due to finite--amplitude perturbations that can experience a significant transient amplification by extracting energy from the laminar flow \citep{Schmid2001}.
Mathematically, transient growth stems from the non-normality of the evolution operator.
The energy gain of a perturbation strongly depends on its shape, $\boldsymbol{u}_0^\prime=\boldsymbol{u}^\prime(r,\theta,z,t=0)$, and, in the nonlinear regime, also on it's initial energy, $E_0=\int|\boldsymbol{u}^\prime_0|^2dV/2$ \citep{Trefethen1993,Schmid1994,Kerswell2018}.
\\
In the linear regime, $E_0 \rightarrow 0$, the optimal perturbation, i.e. the perturbation that experiences the highest energy gain $G=E(\tau)/E_0$ at a time $t=\tau$, consists of a pair of streamwise independent vortices \citep{Schmid1994}. 
This perturbation grows via the lift--up effect to generate large amplitude streaks \citep{EllingsenPalm1975,Landahl1980}. 
Such streaks remain streamwise independent and, regardless of their amplitude, they always decay without triggering transition even when nonlinear effects are taken into account \citep{Zikanov1996}. 
Only by superposing three dimensional residual noise, transition may be triggered by secondary instabilities of the inflectional streaky profile, provided that the initial perturbation amplitude is sufficiently large \citep{Zikanov1996}.
However, \citet{Reddy1998} showed that, in Poiseuille flow, oblique waves are energetically more efficient in triggering transition than two dimensional linear optima with superposed noise.
\\
Various studies investigated critical perturbation amplitudes required to trigger transition as a function of $Re$ for different types of perturbations. 
Generally, the transition threshold was found to follow a power--law scaling, 
\begin{equation}
    E_{0,p}^\text{crit}\propto Re^\gamma,
\end{equation}
with $\gamma=-2.0$ and $\gamma=-3.0$ depending on the details of the perturbation \citep{Hof2003,Meseguer2003_StreakBreakdown,Peixinho2007,Mellibovsky2009}. Similar scaling exponents where found in the asymptotic analysis of \citet{Chapman2002} for different transition routes of streamwise and oblique perturbations in plane Poiseuille flow.
\\
Nonlinear stability analysis enables the computation of the lowest--energy perturbation able to trigger transition, termed the minimal seed, which marks the lowest transition threshold \citep{Pringle2010}.
Nonlinear optimal perturbations (NLOPs) have been computed for boundary--layer flow \citep{Cherubini2010, Cherubini2011}, plane Couette flow \citep{Monokrousos2011, Duguet2010, Duguet2013}, plane Poiseuille flow \citep{Farano2016,Parente2022}, steady pipe flow \citep{Pringle2010,Pringle2012,Pringle2015} or pulsatile pipe low \citep{Keuchel2025}.
In all these systems, NLOPs emerges for sufficiently large $E_0$. 
They are fully localised in space and leverage a combination of Orr and lift--up mechanism connected by a nonlinear oblique phase to significantly outgrow linear optimal perturbations (LOPs) and efficiently trigger transition \citep{Kerswell2018}.
\citet{Marensi2019} computed the critical initial energy of minimal seeds in $L=10R$ long pipes up to large $Re=10^4$ and found a scaling of $E_0^\text{crit} \propto Re^{-2.8}$. They further showed that modifications of the laminar profile can increase the critical energy required to trigger transition, demonstrating how minimal seeds provide a useful framework to quantify susceptibility to transition and ultimately develop control techniques.
Similarly, a scaling of $E_0^\text{crit} \propto Re^{-2.7}$ was found in plane Couette flow \citep{Duguet2013}, however, they noted that their short domain size might effect the scaling exponent. 
The effect of the domain size on the energy growth of NLOPs in pipe flow was studied by \citet{Pringle2015} who showed that accurately capturing the energy growth at $Re=2400$ requires a domain length of about $L\gtrsim50R$.
Further studies of minimal seeds in channel flow, find that an insufficient domain size has a considerable effect on the shape and the critical initial energy of minimal seeds \citep{Parente2022}.
\\
We revisit the scaling of minimal seeds in pipe flow.
In the first part of this paper, \S \ref{sec: The effect of optimisation parameters}, we assess the influence of numerical parameters on the growth of nonlinear optimal perturbations, $G$, and the initial energy of minimal seeds, $E_0^\text{crit}$. These results lay the foundation for a systematic study of the growth mechanisms and the scaling of minimal seeds independent of the numerical parameters.
In the second part of this paper, \S \ref{sec: non linear optimal} and \S \ref{sec: Effect of Re}, we then focus on the effect of the initial energy and Reynolds number on the shape and localisation of NLOPs, their energy growth during different evolution phases and determine the transition threshold. In \S \ref{sec: Discussion}, we compared our results to previous findings and conclude in \S \ref{sec: Conclusion}.

\section{Methods}
\label{sec: Methodology}
We consider the flow of a fluid with constant density $\rho$ and kinematic viscosity $\nu$ through a cylindrical pipe driven with a constant bulk velocity $\bar{U}=2U_c$. We scale all lengths with the pipes radius, $R$, and velocities with the centreline velocity of the laminar Hagen-Poiseuille profile, $U_c$. Considering the laminar Hagen-Poiseuille profile as base flow, we decompose the total velocity field into base flow and perturbation,
\begin{equation}
    \label{eq: flow decomposition}
    \boldsymbol{u}(r,\theta,z,t)=(1-r^2)\boldsymbol{e}_z +\upe(r,\theta,z,t),\hspace{10mm} r\in[0,1]
\end{equation}
and aim to find the optimal initial perturbation $\boldsymbol{u}{'}(r,\theta,z,t=0)=\boldsymbol{u}_0^\prime$, which maximises the energy growth, $G(\tau)={\left\langle |\boldsymbol{u}'(t={\tau})|^2\right\rangle}/{\left\langle|\boldsymbol{u}'_0|^2\right\rangle}={E(t={\tau})}/{E_0}$, at a given time $t=\tau$.
Here $\langle ...\rangle=\int_0^{L} \int_0^{2\pi} \int_0^1  (...)r \ dr  d\theta  dz$ is the volume integral over the pipe of length $L$ and $\tau$ the optimisation time.
Optimal perturbations were computed following the nonlinear variational approach of \citet{Kerswell2014}. 
Specifically, we define the Lagrangian
\begin{eqnarray}
    \mathcal{L}&=&\left\langle\frac{1}{2} |\boldsymbol{u}'(t=\tau)|^2\right\rangle
    \nonumber \\
    && +\lambda \left[  \left\langle\frac{1}{2} |\boldsymbol{u}'_0|^2\right\rangle -E_0\right]
    \nonumber \\
    && +\int_{{0}}^{\tau} \left\langle  \boldsymbol{u}^{\dagger} \boldsymbol{\cdot} \left[ \frac{\partial \boldsymbol{u}'}{\partial t}+(\boldsymbol{U}\boldsymbol{\cdot} {\nabla})\boldsymbol{u}' +(\boldsymbol{u}'\boldsymbol{\cdot} {\nabla})\boldsymbol{U} +(\boldsymbol{u}'\boldsymbol{\cdot} {\nabla})\boldsymbol{u}' +\nabla p' -\frac{1}{Re}\nabla^2\boldsymbol{u}' \right]  \right\rangle \ dt
    \nonumber \\
    && +\int_{{0}}^{{\tau}} \left\langle  p^{\dagger} \left[\nabla \cdot \boldsymbol{u}' \right ]\right\rangle \ dt
    +\int_{{0}}^{{\tau}} \Gamma \left\langle  \boldsymbol{u}' \boldsymbol{\cdot} \boldsymbol{e}_z\right\rangle \ dt,
\end{eqnarray}
where $(\lambda,\boldsymbol{u}^\dagger,p^\dagger,\Gamma)$ are Lagrangian multipliers, known as the adjoint variables, enforcing the initial perturbation energy $E_0$, momentum conservation, mass conservation and a zero bulk velocity of the perturbation to maintain a constant $Re$.
Considering no--slip conditions at the pipe wall and periodicity in streamwise direction, optimal perturbation, i.e. velocity fields for which the first variation of $\mathcal{L}$ vanishes, must satisfy the following equations:
\begin{itemize}
    \item initial energy constraint 
        \begin{eqnarray}
        \frac{\delta \mathcal{L}}{\delta \lambda}= \left[  \left\langle\frac{1}{2} |\boldsymbol{u}'_0|^2\right\rangle -E_0\right]  =0
        \label{eq: energy constraint}
        \end{eqnarray}
    \item direct Navier--Stokes equations
        \begin{equation}
        \frac{\delta \mathcal{L}}{\delta \boldsymbol{u}^{\dagger}}=  \frac{\partial \boldsymbol{u}'}{\partial t}+(\boldsymbol{U}\boldsymbol{\cdot} {\nabla})\boldsymbol{u}' +(\boldsymbol{u}'\boldsymbol{\cdot} {\nabla})\boldsymbol{U} +(\boldsymbol{u}'\boldsymbol{\cdot} {\nabla})\boldsymbol{u}' +\nabla \hat{p}'+f_z \boldsymbol{e}_z -\frac{1}{Re}\nabla^2\boldsymbol{u}'  =\boldsymbol{0}
        \label{eq: direct momentum}
        \end{equation}
        \begin{equation}
        \frac{\delta \mathcal{L}}{\delta p^{\dagger}}= \nabla \cdot \boldsymbol{u}' =0
        \label{eq: direct continuity}
        \end{equation}
    \item zero bulk velocity of the direct perturbation field
        \begin{equation}
        \frac{\delta \mathcal{L}}{\delta \Gamma}= \left\langle  \boldsymbol{u}' \boldsymbol{\cdot} \boldsymbol{e}_z\right\rangle   =0
        \label{eq: direct flux constraint}
        \end{equation}
    \item compatibility condition
        \begin{eqnarray}
        \frac{\delta \mathcal{L}}{\delta \boldsymbol{u}'(t={\tau})}= \boldsymbol{u}^\prime(t={\tau})+\boldsymbol{u}^{\dagger}(t={\tau})  = \boldsymbol{0}
        \label{eq: compatibility}
        \end{eqnarray}
    \item adjoint Navier--Stokes equations
        \begin{equation}
            \frac{\delta \mathcal{L}}{\delta \boldsymbol{u}'}=-\frac{\partial \boldsymbol{u}^{\dagger}}{\partial t}-\left[(\boldsymbol{U}+\boldsymbol{u}')\cdot \nabla\right]\uad+\left[\nabla (\boldsymbol{U}+\boldsymbol{u}')\right]^T\uad -\nabla \hat{p}^{\dagger}+f_z^{\dagger}\boldsymbol{e}_z-\frac{1}{Re}\nabla^2\uad =\boldsymbol{0}
            \label{eq: adjoint momentum}
        \end{equation}
        \begin{equation}
            \frac{\delta \mathcal{L}}{\delta \hat{p}'}=   \nabla \cdot \uad=0
            \label{eq: adjoint continuity}
        \end{equation}
    \item zero bulk velocity of the adjoint perturbation field
        \begin{equation}
            \frac{\delta \mathcal{L}}{\delta f_z}= \left\langle  \uad \boldsymbol{\cdot} \boldsymbol{e}_z\right\rangle     =0
            \label{eq: adjoint flux constraint}
        \end{equation}
    \item optimality condition
        \begin{eqnarray}
        \frac{\delta \mathcal{L}}{\delta \boldsymbol{u}'_0}= \lambda \upe_0 -\uad_0   =\boldsymbol{0}.
        \label{eq: optimality}
        \end{eqnarray}
\end{itemize}
To satisfy equations \eqref{eq: energy constraint}--\eqref{eq: optimality}, we employ the classical adjoint looping procedure \citep{Kerswell2018} in which an initial perturbation guess, with initial energy $E_0$, is integrated forward in time, $0\leq t \leq \tau$, via the direct Navier--Stokes equations \eqref{eq: direct momentum}--\eqref{eq: direct flux constraint} with a zero perturbation flux constraint. With the solution of the direct field at time $t=\tau$, the adjoint velocity field is initialised using equation \eqref{eq: compatibility} and then integrated backward in time, $t=\tau$ to $t=0$, via the adjoint Navier--Stokes equations \eqref{eq: adjoint momentum}--\eqref{eq: adjoint flux constraint} with same boundary conditions and flux constraint as the direct field.
With the direct and adjoint field at $t=0$ the gradient $(\delta \mathcal{L}/\delta\boldsymbol{u}_0)$ is computed. Generally, the optimality condition is not satisfied by the initial guess, $(\delta \mathcal{L}/\delta\boldsymbol{u}_0)\neq 0$, however, it enables the iterative optimisation of the initial condition $\boldsymbol{u}_0^\prime$, e.g. with gradient ascent method.
\\
The direct--adjoint optimisation procedure was implemented in the GPU version of the open source pseudo--spectral Navier--Stokes code \href{https://github.com/Mordered/nsPipe-GPU}{\texttt{nsPipe-GPU}} \citep{Lopez2020,Moron2024}. The direct and adjoint equations are discretised by a Fourier--Galerkin method in the homogenous directions and sixth--order finite--differences in the radial direction. Time integration is performed with a second--order predictor corrector method with a fixed time step $\Delta t$ and a flux correction during the corrector step to maintain a zero bulk velocity of the perturbation field. A logarithmic checkpointing was employed to reduce the time of recovering direct fields during the integration of the adjoint equations \citep{Keuchel2025}. The initial perturbation was iteratively moved towards an optimum using a gradient ascent method with a backtracking line search for the step size.
For details of the numerical solution procedure, optimisation step, checkpointing procedure and a code validation, we refer the reader to \citet{Keuchel2025}. 

\section{The effect of optimisation parameters}
\label{sec: The effect of optimisation parameters}
\subsection{Convergence and robustness of the procedure to the initial condition}
\begin{figure}[h]
    \centering
    \includegraphics[width=0.99\linewidth]{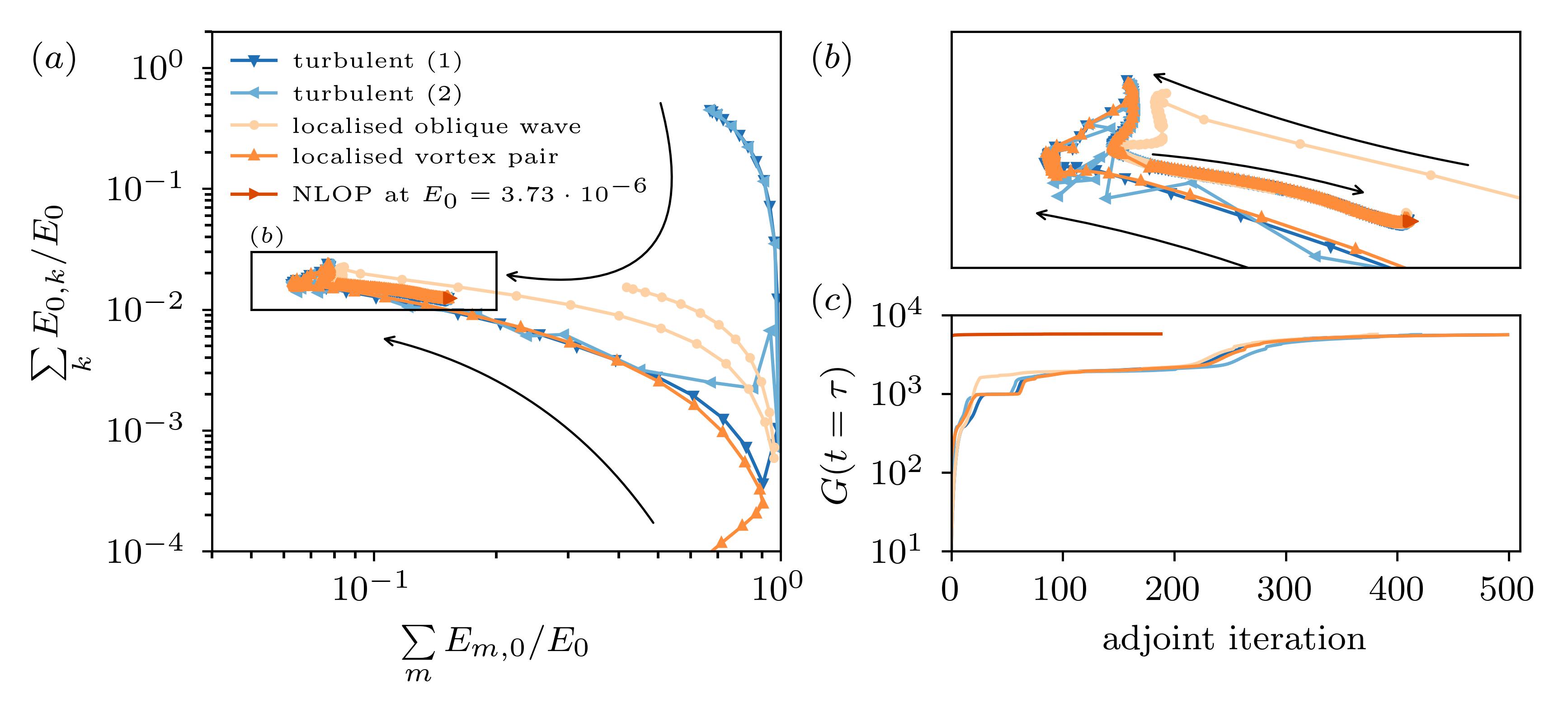}
    \caption{$(a)$ Convergence of the optimisation procedure initialised with different initial guesses and $(Re,E_0,L,\tau)=(2400,3.7\cdot 10^{-5},10,150)$. Each marker indicates the modal energy of the $m=0/k=0$ azimuthal/axial modes, normalised by $E_0$, of the improved initial perturbation at each adjoint iteration. $(b)$ A magnification of $(a)$ close to convergence. $(c)$ The energy gain of the improved initial perturbation achieved at the optimisation time as a function of the adjoint iterations.
    }
    \label{fig: convergence study}
\end{figure}

In general, the nonlinear optimisation problem is non--convex and may features multiple local optima. In order to verify the robustness of the optimisation procedure, we computed optimal perturbations for fixed parameters, $(Re,E_0,L)$, initialising the optimisation with different perturbation guesses. Specifically, NLOPs were computed at $(Re,E_0,L,\tau)=(2400,3.7\cdot 10^{-5},10,150)$. Optimisations were initialised with 
\begin{enumerate}[label=\roman*), leftmargin=*, align=left]
    \item A rescaled turbulent field 
    \item A streamwise localised pair of counter rotating vortices
    \item A streamwise localised, oblique wave obtained from the evolution of an NLOP
    \item A NLOP computed at a different initial energy.
\end{enumerate}
In figure \ref{fig: convergence study}$a$, we show the evolution of the improved initial perturbations in phase space at each adjoint iteration (each marker indicates the next adjoint iteration) together with the corresponding energy gain as a function of the adjoint iteration, figure \ref{fig: convergence study}$c$.
Depending on the initial guess, the optimisation procedure converges after $200-500$ adjoint iterations. Independent of the initial guess, however, the optimisation converges to the same optimal perturbation, with a relative error of the energy gain, $G(t=\tau)$, between $1\%-2\%$ for the different initial guesses.
This error could be reduced by setting a smaller tolerance of the relative improvement of the energy gain between adjoint iterations. 
However, this significantly increases the computational cost due to the additional adjoint iterations.
Here, the optimisation was terminated once the relative improvement between adjoint iterations $(s)$, drops below a tolerance, $(G^s-G^{s-1})/G^s<10^{-5}$, as a compromise between accuracy and computational cost.

\subsection{Optimisation time}
\label{ssec: optimisation time}
\begin{figure}[h]
    \centering
    \includegraphics[width=0.99\linewidth]{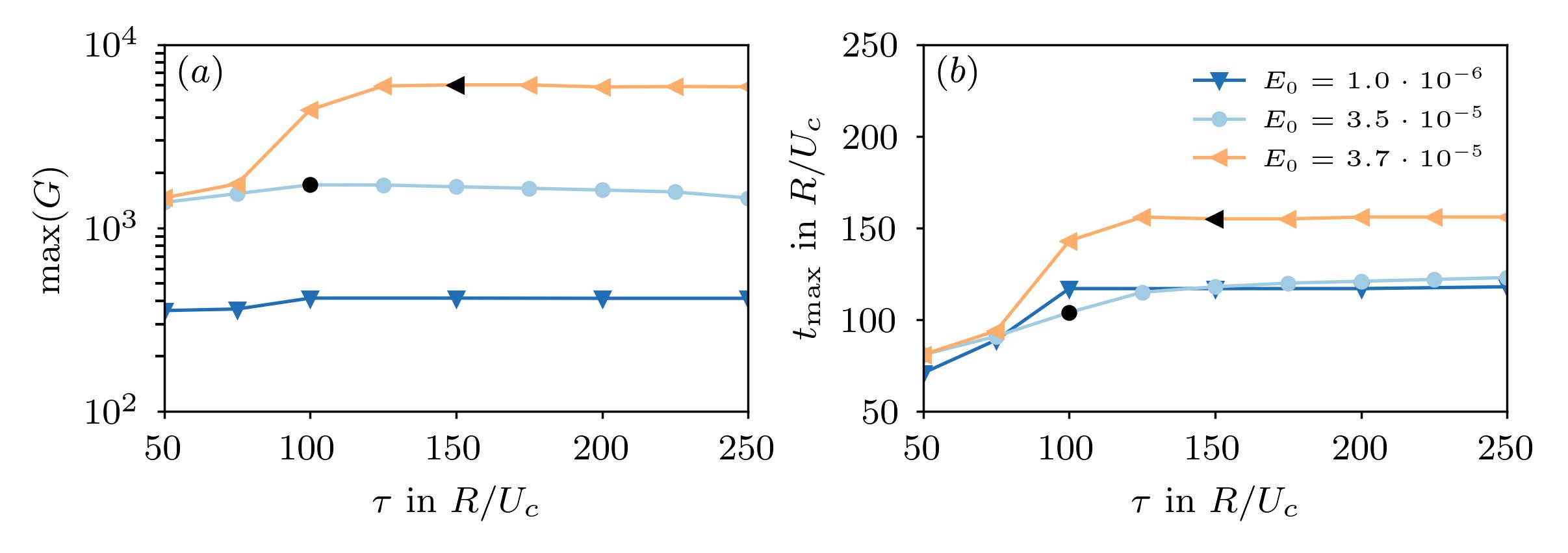}
    \caption{$(a)$ The maximum energy, $\max (G)$, and $(b)$ time of maximum amplification $t_\text{max}=\arg \max (G)$ of optimal perturbations of different initial energy, $E_0$, as a function of the optimisation time $\tau$ for $(Re,L)=(2400,10)$. The black marker indicates the optimisation time leading to the NLOP with highest gain and the black dotted line in $(b)$ corresponds to $\tau=t_\text{max}$ as an indicator on how good the choice of $\tau$ is.
    }
    \label{fig: optimisation time study}
\end{figure}

To study the dependence of optimisation results on the optimisation time, $\tau$, optimal perturbations were computed at $(Re,L)=(2400,10)$. In figure \ref{fig: optimisation time study}, we show the maximum energy gain and time of maximum amplification as a function of $\tau$ for different initial energies, $E_0$.
For a sufficiently small initial energy, $E_0=10^{-6}$, the optimisation procedure converges to the linear optimal perturbation provided $\tau \gtrsim 100$ (dark blue line in figure \ref{fig: optimisation time study}$a$). The linear optimal perturbation experiences a maximum growth of $\max(G)=415$ at $t_\text{max}=117$, in quantitative agreement with \citet{Meseguer2003}. Only for short optimisation times, $\tau \lesssim 75$, the optimisation procedure converges to a perturbation consisting of two pairs of streamwise independent vortices, which experiences a higher energy growth at shorter times, but a lower overall energy gain. 
For higher initial energies, $E_0=3.5\cdot 10^{-5}$ and $E_0=3.7\cdot 10^{-5}$, nonlinear optimal perturbations emerge and achieve a significantly higher energy gain than the linear optimal perturbation.
Similarly to the linear optimal perturbation, optima for short optimisation times grow quickly but experience a lower maximum amplification.
For sufficiently long $\tau$, which increases with $E_0$, optimisation results are robust w.r.t. $\tau$ (figure \ref{fig: optimisation time study}). Especially close to the minimal seed, here $E_0^\text{crit}\approx3.73 \cdot 10^{-5}$, the maximum energy gain of optima is almost independent of $\tau$ provided that $\tau\gtrsim 125$, whereby the maximum relative deviation of the energy gain is about $2.5\%$.
Similarly to the maximum energy gain, the time of maximum amplification, $t_\text{max}=\arg \max(G)$, is robust to the choice of $\tau$ (figure \ref{fig: optimisation time study}$b$). Hence, $t_\text{max}$ can be used as an indicator to improve the choice of $\tau$.
Specifically, throughout this paper, $\tau$ was initially set to the time of the linear optima, and adjusted if $\tau\ll t_\text{max}$ or $\tau\gg t_\text{max}$, where $t_\text{max}$ was based on previous optimisation results. 
Close to the minimal seed we found that $t_\text{max}^\text{NLOP}\approx t_\text{max}^\text{LOP}$ and that NLOPs are quite robust to the choice of $\tau$, in agreement with \citet{Kerswell2018}.
\begin{figure}[h]
    \centering
    \includegraphics[width=0.99\linewidth]{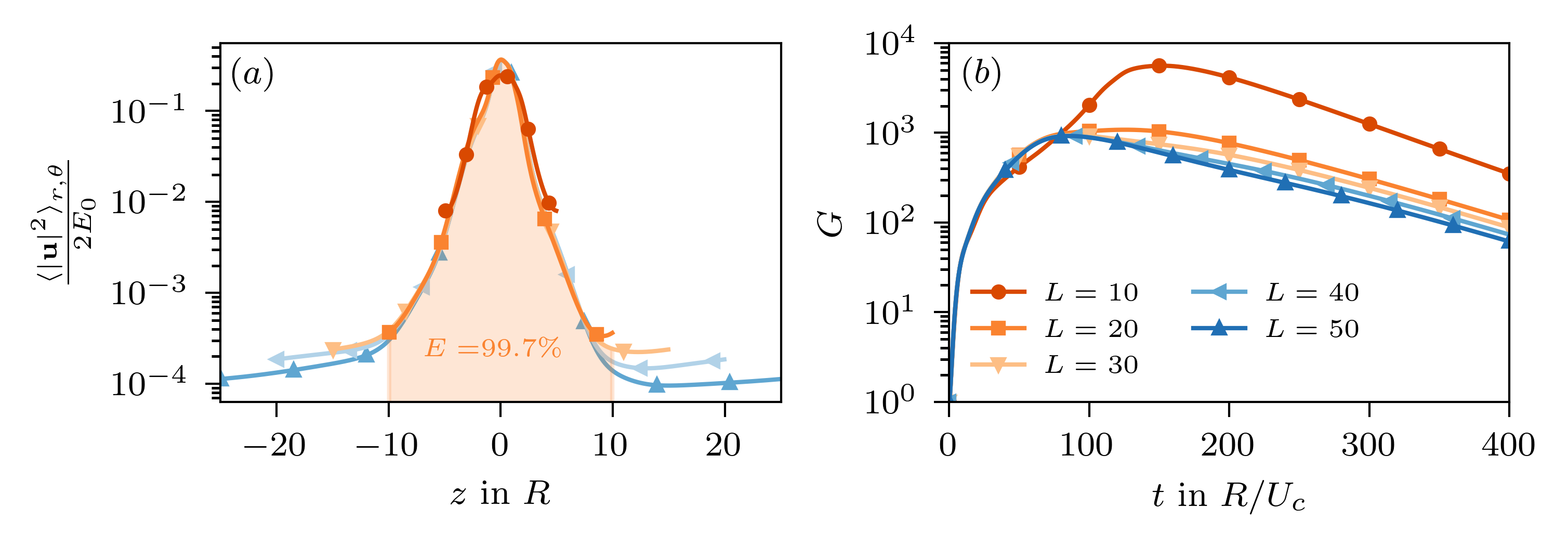}
    \caption{$(a)$ The normalised energy density, $e_{r,\theta}(z)=\langle |\boldsymbol{u}_0|^2\rangle_{r,\theta}/(2E_0 )$, of nonlinear optimal perturbation computed for different pipe length, $L$, and $(Re,E_0)=(2400,3.7\cdot 10^{-5})$. The shaded area indicates the region that contains $99.7\%$ of the initial energy $E_0$.
    $(b)$ The energy gain over time of the different NLOPs shown in $(a)$.}
    \label{fig: length study}
\end{figure}

\subsection{Computational domain length}
\label{ssec: Computational domain length}
We performed  optimisations at $(Re,E_0)=(2400,3.7\cdot 10^{-5})$ and $L\in[10,50]$ to determine the necessary pipe length for the energy gain to be unaffected by the periodic boundary conditions in the streamwise direction.
Figure \ref{fig: length study}$a$, shows the relative energy density,
\begin{equation}
    \frac{e_{r,\theta}(z)}{E_0}=\frac{1}{2E_0}\int _0^{2\pi} \int_0^1 |\boldsymbol{u}|^2r \ drd\theta,
\end{equation}
as a function of the streamwise coordinate $z$ of the localised NLOPs computed for each $L$. Qualitatively, the NLOP is less localised for $L=10$, but does not change substantially for pipe length $L\geq20$. Although $99.7\%$ of the initial energy are contained in a $20R$ long section of the pipe, during its evolution, the perturbation delocalises which affects the energy even for $L\geq20R$ (see figure \ref{fig: length study}$b$).
We find that the time of maximum amplification and the corresponding energy gain are unaffected for $L\geq40$, and we set $L=50$.
\\
We computed the critical initial energy, $E_0^\text{crit}$, for $L=10$ and $L=50$ at $Re=4000$. In the short pipe setup, we find $8.25 \cdot 10^{-5}<E_0^\text{crit}<8.5 \cdot 10^{-5}$, whereas for long pipes, $9.35 \cdot 10^{-5}<E_0^\text{crit}<9.4 \cdot 10^{-5}$. This corresponds to a relative error of about $10\%$ and reveals that a pipe length of $L=50$ is required to accurately resolve both $G$ and $E_0^\text{crit}$ for $Re\leq4000$.
By varying the optimisation time, $\tau\in[150,200,250]$, for the longer pipe setup, $(Re,L,E_0)=(4000,50,9.35 \cdot 10^{-6})$, we find a smaller energy gain $G$ of about $22\%$ if $\tau$ is too short and only a small difference of about $2\%$ for $\tau$ (as discussed in \S \ref{ssec: optimisation time}). However, for all $\tau$ considered, the critical initial energy is $9.35 \cdot 10^{-6}<E_0^\text{crit}<9.4 \cdot 10^{-6}$ which corresponds to relative deviation of less than $0.5\%$. Compared to the energy gain, the transition threshold is less sensitive to the choice of $\tau$.
\begin{figure}[h]
    \centering
    \includegraphics[width=0.99\linewidth]{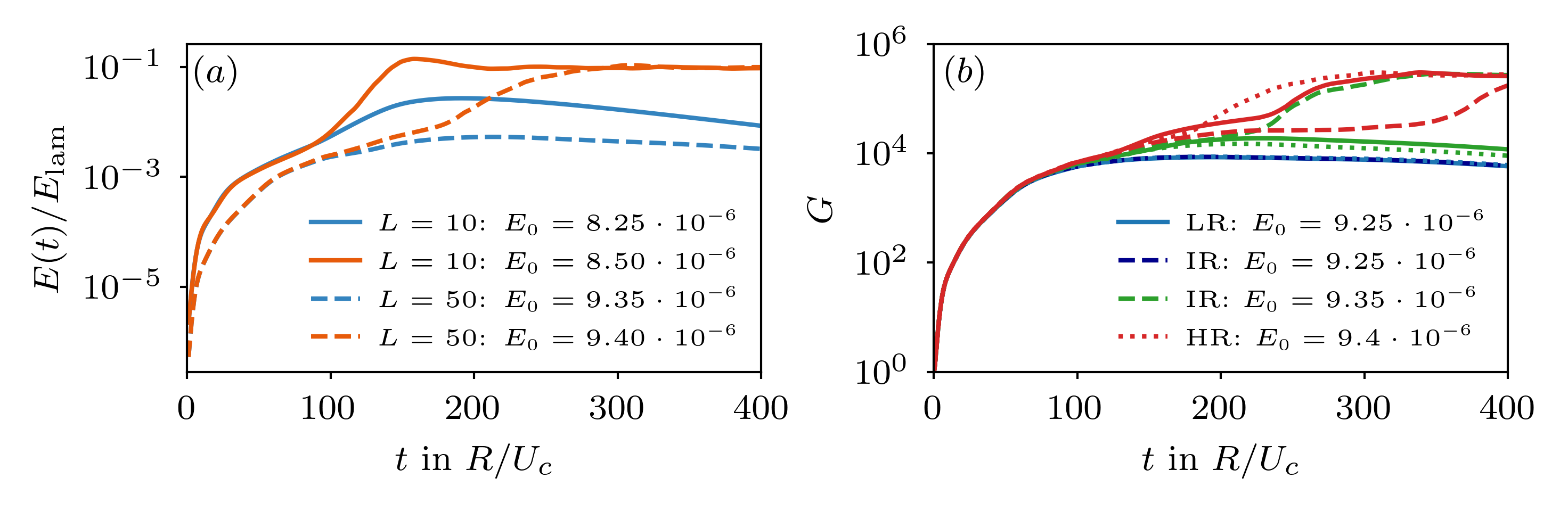}
    \caption{$(a)$ The perturbation energy over time normalised by the laminar energy, $E_\text{lam}=\pi L/6$, of NLOPs with $E_0$ below and above the minimal seed for two pipe length $L=10$ and $L=50$. 
    $(b)$ The energy gain over time of NLOPs computed for different initial energies $E_0$ with three different numerical resolutions. LR: $(N_r,N_\theta,N_z,\Delta t)=(56,48,218,0.005)$ (solid line). IR: $(N_r,N_\theta,N_z,\Delta t)=(76,64,290,0.0025)$ (dashed line). HR: $(N_r,N_\theta,N_z,\Delta t)=(96,80,360,0.00125)$ (dotted line).}
    \label{fig: sensitivity study}
\end{figure}

\subsection{Numerical resolution}
\label{ssec: numerical resolution}
We determine the necessary numerical resolution in $L=50$ long pipes. As a reference resolution, the number of radial points, azimuthal modes and axial modes was set to $(N_r,N_\theta,N_z)=(96,80,360)$, with a time step of $\Delta t\in[5\cdot 10^{-3},5\cdot 10^{-4}]$, depending on the Reynolds number. The spatial resolution was verified by performing a simulation of the last optimal perturbation inside the laminar attractor for $Re=4000$ with $(N_r,N_\theta,N_z,\Delta t)=(144,120,540,0.00125)$. We found that the error in the energy gain is about $0.3\%$ which was deemed sufficiently accurate.
We then successively reduce the spatial resolution to $(76,64,290)$ and $(56,48,218)$ and double the time step, $\Delta t=0.0025$ and $\Delta t=0.005$ respectively. For the different resolutions, we computed NLOPs for three different initial energies, shown in figure \ref{fig: sensitivity study}$b$, and compute the relative error in the energy gain $\epsilon=(G_\text{max}-G_\text{max}^\text{ref})/G_\text{max}^\text{ref}$, where $G_\text{max}$ is the maximum energy gain of a reduced resolution and $G_\text{max}^\text{ref}$, the maximum energy gain with the reference resolution.
For sufficiently low initial energy, $E_0=9.25 \cdot 10^{-6}$, the energy gain is almost independent of the resolution, $\epsilon \lesssim 2\%$. Close to the minimal seed, $E_0=9.35 \cdot 10^{-6}$, the error increases significantly with $\epsilon \approx 26\%$ for the lower resolution and an NLOP above the minimal seed for the intermediate resolution. Despite the large error in $G$, however, the relative error of the critical initial energy, remains below $1.6\%$ for the three resolutions considered. 
Hence, the transition threshold can be accurately computed with a much coarser discretisation than the energy gain.
\\
If not stated otherwise, optimal perturbations were computed with the reference resolution, $(N_r,N_\theta,N_z)=(96,80,360)$, in $L=50$ long pipes sufficient to resolve both $G$ and $E_0^\text{crit}$.

\section{Optimal perturbations}
\label{sec: non linear optimal}
\begin{figure}[h]
    \centering
    \includegraphics[width=0.99\linewidth]{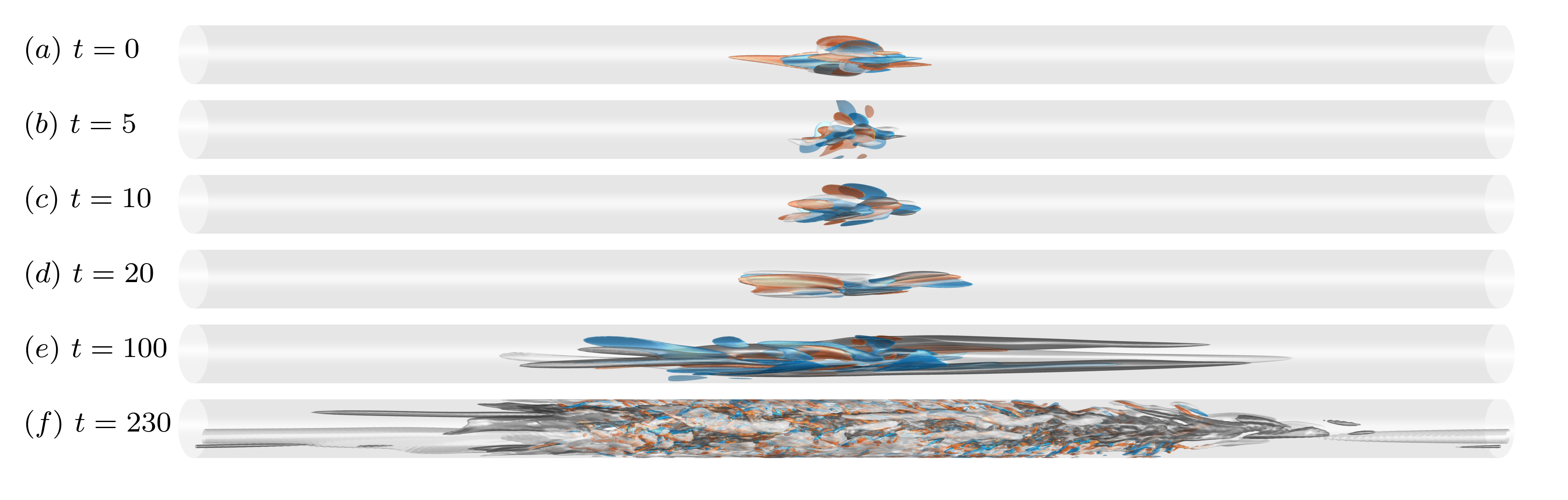}
    \caption{Iso-contours of streamwise velocity ($20\%$ of $\max(u_z)$ in white and $\min(u_z)$ in black) and vorticity ($10\%$ of $\max(\omega_z)$ in orange and $\min(\omega_z)$ in blue) at selected times for $(Re,E_0)=(4000,9.4\cdot 10^{-6})$.
    }
    \label{fig: NLOP evolution turbulent}
\end{figure}

With the verified numerical setup, we now focus on the effect of $(E_0,Re)$ on optimal perturbations.
We first consider a constant Reynolds number, $Re=4000$, and analyse the evolution of optimal perturbations at various $E_0$.
\\
For small initial energies, $E_0 \lesssim 5.5 \cdot 10^{-6}$, the optimum corresponds to the LOP reaching a maximum energy growth of $G=1152$ at $t=195$, in agreement with \citet{Meseguer2003}. 
For sufficiently large initial energies, $E_0 \gtrsim 7.5 \cdot10^{-6}$, nonlinear effects become substantial and a NLOP emerges.
In the nonlinear regime, the energy gain $G$ increases significantly with $E_0$ and, for $E_0^\text{crit} \approx 9.4\cdot 10^{-6}$, the NLOPs trigger transition in form of a turbulent slug (figure \ref{fig: NLOP evolution turbulent}).
Qualitatively, NLOPs are identical at all $E_0$ considered and leverage the same growth mechanisms. 
In the following, we discuss the evolution phases of NLOPs on the basis of $E_0=9.35\cdot 10^{-6}$ and point out quantitative differences between different $E_0$.

\subsection{The Orr phase}
\begin{figure}[h]
    \centering
    \includegraphics[width=0.99\linewidth]{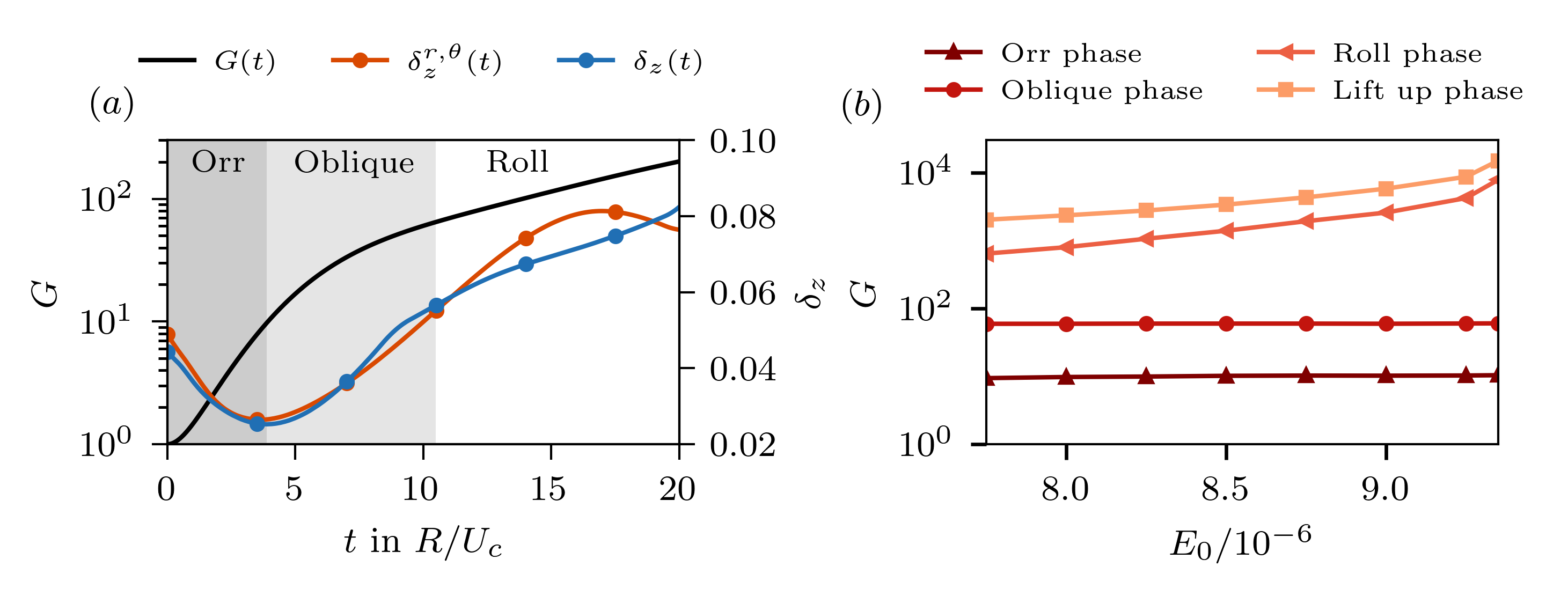}
    \caption{
    $(a)$ The energy gain $G$, axial thickness of rolls, $\delta^{(r,\theta)}_z$, and axial thickness, $\delta_z$ as a function of time in the perturbations early evolution stage, $t\leq20$, for $Re=4000$ and $E_0=9.35\cdot10^{-6}$.
    $(b)$ The energy gain at the end of the Orr phase, oblique phase, roll phase and lift--up phase for NLOPs at various $E_0$ and $Re=4000$.
    }
    \label{fig: delta and G(E0)}
\end{figure}

Initially, the NLOP is characterised by three dimensional velocity layers that lean against the base shear. During the perturbation's evolution, these layers are tilted until they align with the base shear (see figure \ref{fig: NLOP evolution turbulent}$a,b$).
As shown in figure \ref{fig: delta and G(E0)}$a$, this reorientation leads to a short but rapid energy growth. Most of the energy remains in the initially excited Fourier mode of large streamwise wavenumber, $k>12$.
This is reminiscent of the Orr mechanism \citep{Orr1907, Jimenez2013,Chagelishvili2016}.
To study the energy gain during the Orr phase, we determine the time, $t_\text{Orr}$, at which the relative axial thickness of the localised perturbation,
\begin{equation}
    \label{eq: axial thickness}
    \delta_z(t)=
    \frac{1}{L}  \int \limits_0^{L}  
    \frac{e_{r,\theta}(z)}
    {\max \limits_z \left( e_{r,\theta}(z)\right)}
     dz,
\end{equation}
attains a minimum (figure \ref{fig: delta and G(E0)}$a$),
\begin{equation}
    t_\text{Orr}=\arg \min(\delta_z(t)) \hspace{10mm} \text{for }t<20.
\end{equation}
Hence, $\delta=1$ indicates a field that is independent of the spatial coordinate, while $\delta\approx 0$ corresponds to a Dirac impulse.
We found, that this corresponds to the time, $t_\text{Orr}\approx 4.5$, where the localised, and initially tilted velocity layers, are normal to the streamwise direction, i.e. to the time of maximum Orr amplification.
In figure \ref{fig: delta and G(E0)}$b$, we show that the energy amplification during the initial Orr phase, $G_\text{Orr}=G(t_\text{Orr})$, is independent of $E_0$ indicating that nonlinearities do not play a substantial role, as expected from the essentially linear mechanism and small perturbation energy at this stage. 

\subsection{Oblique--wave phase}
Once the perturbation aligns with the base flow, $t\approx 4.5$, it further gains energy and rapidly elongates.
This phase was previously termed the \textit{oblique--wave phase} and ends once the energy of $k\neq0$ modes reach a maximum \citep{Pringle2010,Pringle2012}. For their considered pipe length of $L=\pi$, oblique modes feed energy into streamwise independent rolls via nonlinear interaction reflected by an increase of $E_{r,\theta}^{(m,0)}$ (see figure 3 of \citet{Pringle2012}).
For longer pipes, however, this initial elongation of rolls does not correspond to a growth of $k=0$ modes, but rather to an energy transfer into oblique modes of lower streamwise wavenumber. In physical space, rolls elongate, but due to the longer domain, they remain localised. This was recognised by \citet{Marensi2019}, who studied minimal seeds for $L=10$ and defined the oblique phase as the time of maximum amplification of oblique modes with streamwise wavenumber $k>3$.
Analogously we could specify a streamwise wavenumber threshold to define the oblique phase for $L=50$.
However, since the physical extent of the NLOP remains the same for $L\geq20$ (as shown in figure \ref{fig: length study}$a$), the appropriate wavenumber $k$ depends on the pipe length.
To define the oblique phase, we propose a physical definition based on the characteristic delocalisation of the rolls:
We define the end of the oblique phase (or \textit{initial elongation phase}) as
\begin{equation}
    t_\text{obl}=\arg \max \left( \frac{d\delta_z^{(r,\theta)}}{dt}\right),
\end{equation}
where $\delta_z^{(r,\theta)}$ is the streamwise thickness based on the energy destiny of the cross sectional velocity components, $(u_t,u_\theta)$, i.e. the time of maximum elongation rate of rolls (here $t_\text{obl}\approx10.5$), which is independent of the domain length (not shown). 
As the relative axial thickness is normalised by the instantaneous maximum of the energy density, eq. \eqref{eq: axial thickness}, we found that its temporal derivative can be noisy. To obtain a smooth derivative and well defined $t_\text{obl}$, we normalise $\delta_z^{(r,\theta)}(t)$ by the moving average of the maximum energy density rather than the instantaneous maximum energy density. (we average the maximum energy density over a time of $1R/U_c$, however, halving or doubling this time does not significantly affect the result). 
\\
In figure \ref{fig: delta and G(E0)}$b$, we show the energy gain of NLOPs with different $E_0$ during the oblique phase. Despite the nonlinear energy transfer between oblique modes of different $k$, the energy gain is almost independent of $E_0$. This suggests that nonlinear effects play only a secondary role in the perturbation amplification during the oblique phase.
\begin{figure}[h]
    \centering
    \includegraphics[width=0.99\linewidth]{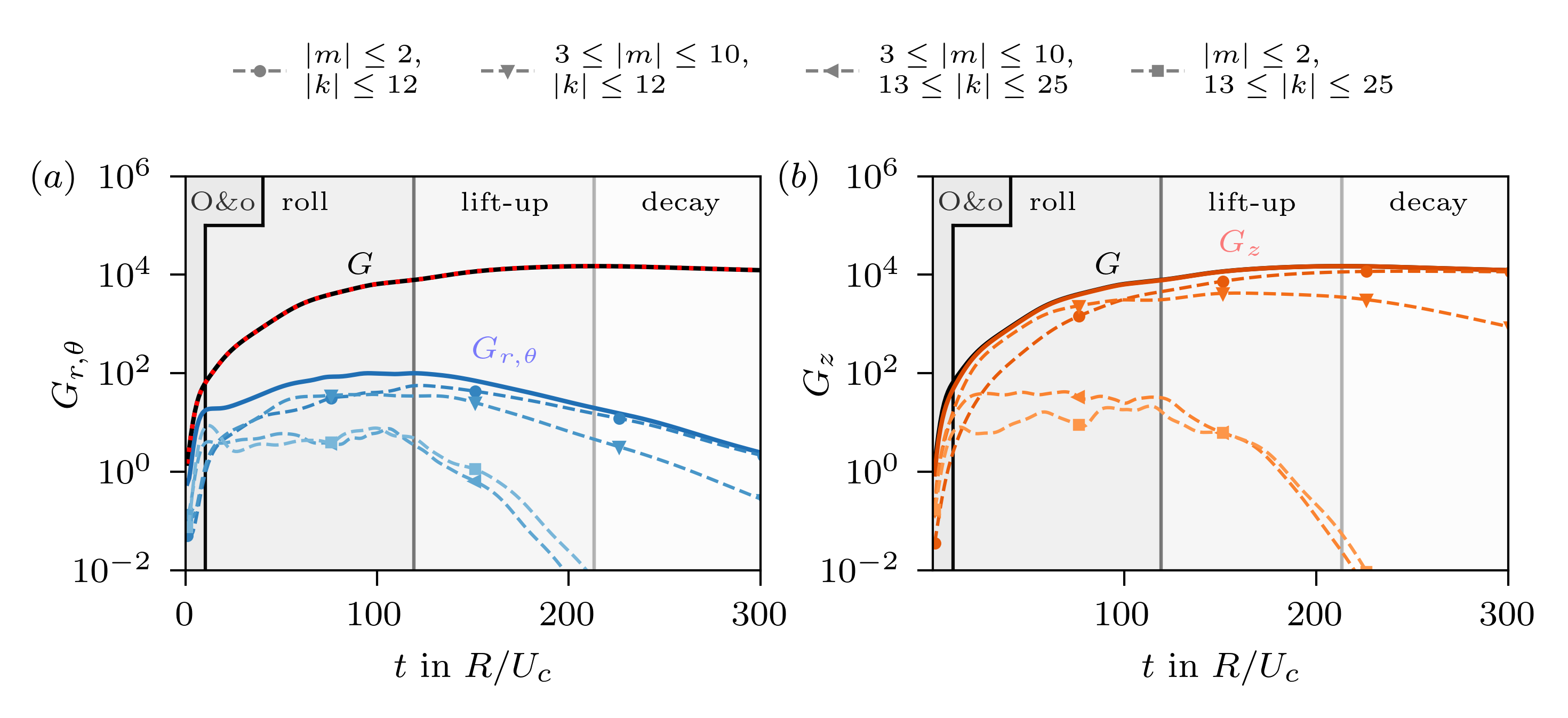}
    \caption{
    $(a)$ The total energy gain $G$, the energy gain of the cross-sectional components $G_{r,\theta}$ and the energy gain of the cross-sectional components of selected azimuthal modes, $m$, and axial modes $k$, $\sum_{m}\sum_{k} G_{r,\theta}^{(m,k)}$ over time. 
    Shaded areas indicate the Orr and oblique phase (O\&o), the roll phase and the lift--up phase.
    $(b)$ The same as $(a)$, but for the streamwise component. 
    }
    \label{fig: NLOP}
\end{figure}

\subsection{Nonlinear roll phase}
\begin{figure}[h]
    \centering
    \includegraphics[width=0.99\linewidth]{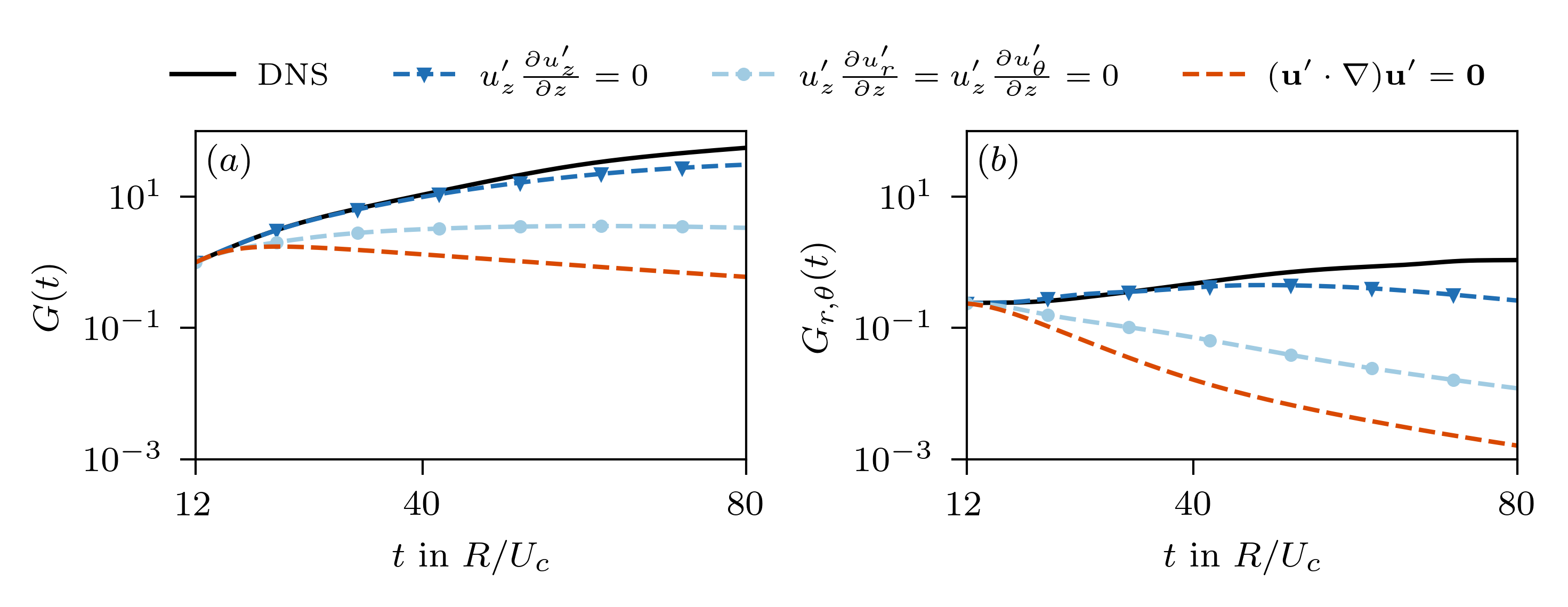}
    \caption{
    $(a)$ The total energy gain and $(b)$ the cross-sectional component of the energy gain of the NLOP for $Re=4000$, $E_0=9.35\cdot 10^{-6}$ over time for $t>12$ where $G(t)=E(t)/E(t=12)$. The dashed line corresponds to the linearised simulation and the solid line to the fully nonlinear simulation. Semi-nonlinear simulations are indicated by markers.
    }
    \label{fig: origin of rolls}
\end{figure}

After the rapid initial elongation of the perturbation, both the energy in rolls (cross-sectional velocities) and streaks increases by $\mathcal{O}(G_{r,\theta})=\mathcal{O}(E_{r,\theta}/E_0)\approx 10^1$ and $\mathcal{O}(G_z)=\mathcal{O}(E_{z}/E_0)\approx 10^3$ respectively (figure \ref{fig: NLOP}). 
Streaks grow by extracting energy from the base flow via the lift--up mechanism, i.e. radial convection of the base shear by the cross--flow.
The growth of rolls stems from the interaction of oblique waves \citep{Reddy1998}.
To further clarify the origin of the roll growth, we perform simulations in which we suppress various nonlinear convective terms and observe how this affects the perturbation's evolution after the Orr and oblique--wave phase. 
In figure \ref{fig: origin of rolls}, we show the growth of the fully nonlinear, semi-nonlinear and linear evolutions of the perturbation for $t>12$.
During the linear evolution, $(\boldsymbol{u}^\prime\cdot\nabla)\boldsymbol{u}^\prime=0$, the perturbation still experiences a short term energy gain, but rolls decay monotonously.
By solely suppressing the streamwise convection of cross sectional components, $u_z^\prime \partial  u_r^\prime/\partial z=u_z^\prime \partial  u_\theta^\prime/\partial z=0$, the perturbation is able to grow for a longer time, leading to a slightly increased energy gain, but rolls also decay monotonously. 
In order for rolls to grow, the perturbation relies on the streamwise convection of cross-flow velocity gradients, $u_z^\prime \partial u_r^\prime /\partial z$ and $u_z^\prime \partial u_\theta^\prime/\partial z$.
At finite perturbation amplitude, substantial nonlinear streamwise advection, $u_z ^\prime\partial \boldsymbol{u}^\prime/\partial z$, of NLOPs amplifies streamwise gradients via self advection. 
Specifically, $u_z^\prime \partial u_r^\prime /\partial z$ and $u_z^\prime \partial u_\theta^\prime/\partial z$ locally amplify $\partial u_r^\prime/\partial z$ and $\partial u_\theta^\prime/\partial z$, which corresponds to an increase of radial and azimuthal vorticity $(\omega^\prime_r, \omega^\prime_\theta)$. 
The components $(\omega^\prime_r, \omega^\prime_\theta)$ are then tilted into the streamwise vorticity component via $(\boldsymbol{\omega}^\prime\cdot \nabla)u_z^\prime$, which explains the growth of rolls and its dependence on $u_z^\prime \partial u_r^\prime /\partial z$ and $u_z^\prime \partial u_\theta^\prime/\partial z$.
We note that for streamwise independent velocity fields, $\partial\boldsymbol{u}^\prime/\partial z=0$, the tilting/stretching terms in the streamwise vorticity equation vanish, $(\boldsymbol{\omega}^\prime\cdot \nabla)u_z^\prime=0$, which inhibits the growth of streamwise rolls. 
In summary, the rolls rely on the waviness of large amplitude streaks.
Simultaneously, the nonlinear interaction of oblique modes causes an energy transfers between different modes of higher and lower wavenumber. Viscous dissipation dominates in large wavenumber modes and consequently, the relative energetic composition is quickly dominated by modes of low wavenumber (see the increasing energy of $k\leq12$ in figure \ref{fig: NLOP}$b,c$). In physical space, this is reflected by a further delocalisation of the perturbation. In figure \ref{fig: delta and G(E0)}$b$, we show the perturbation's energy gain during the roll phase, $G(t_\text{roll})$, where 
\begin{equation}
    t_\text{roll}=\arg \max (G_{r,\theta}(t)).
\end{equation}
During this phase, $G$ increases significantly with $E_0$, underlining the substantial effect of nonlinearities.

\subsection{Lift--up phase}
At around $t\approx 122$, the delocalisation and elongation of the perturbation and the corresponding reduction of streamwise gradients, inhibits the growth of rolls. Thereafter vortices decay monotonously and the perturbation grows solely via the lift--up effect (figure \ref{fig: NLOP}). 
Generally, various oblique modes grow during this stage, whereas the energy gain of each mode strongly depends on $(m,k)$. By nonlinear mixing, the perturbation gradually transfers energy from modes that grow in the short term to modes that grow in the long term. Specifically large wavenumber oblique modes tend to gain more energy over short times but achieve a lower maximum amplification than low wavenumber modes \citep{Schmid1992,Schmid1994,Meseguer2003}. Hence, by transferring energy to the large scales, the optimal perturbation leverages the growth of various oblique modes on different time scales to optimise the amplification. 
\\
Eventually, the amplitude of rolls is insufficient to further amplify the streaks, $t_\text{lift}=\arg \max(G(t))$, and these decay viscously while the continued energy transfer into small wavenumber modes leads to a further delocalisation.
For $E_0 \geq 9.4\cdot 10^{-6}$, the streaks become unstable, break down and subsequently trigger transition in form of a turbulent slug, marking the minimal seed at $Re=4000$.
\\
Although the net energy gain must stem from linear processes \citep{Henningson1996,Schmid2001}, the growth during the oblique--wave and lift--up phase is enhanced by nonlinearities \citep{Kerswell2014, Kerswell2018} and increases significantly with $E_0$ (figure \ref{fig: delta and G(E0)}$b$).
As the growth of rolls depends on $u_z^\prime\partial \boldsymbol{u}^\prime/\partial z$, their time of growth and energy gain increase nonlinearly with $E_0$.
The stronger and longer sustained rolls, lead to a prolonged lift--up phase and consequently to a further increase of $G$ with $E_0$. 

\section{The effect of Reynolds number}
\label{sec: Effect of Re}
In this section, we study the effect of $Re$ on the structure of the minimal seed, the energy gain of NLOPs during each of the previously discussed phases and finally determine the transition threshold.

\subsection{Energy gain along growth phases}
\begin{figure}[h]
    \centering
    \includegraphics[width=0.99\linewidth]{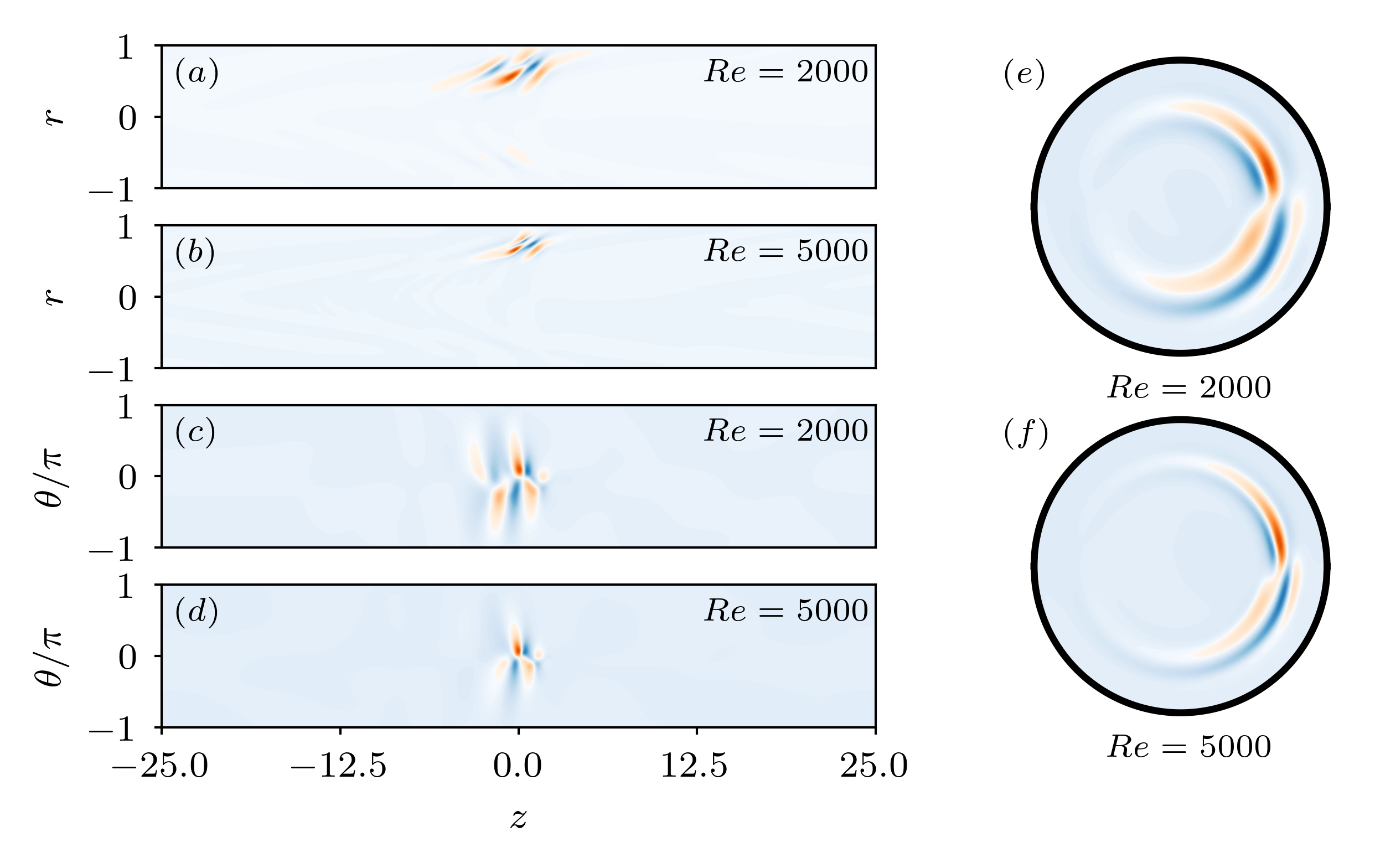}
    \caption{Snapshots of the streamwise velocity, $u_z$, of the NLOP at $Re=2000$ and $Re=5000$ in selected planes at a constant $\theta$, $r$ and $z$ corresponding to the position of the respective maximum energy density.
    }
    \label{fig: ms localisation snapshots}
\end{figure}

\begin{figure}[h]
    \centering
    \includegraphics[width=0.99\linewidth]{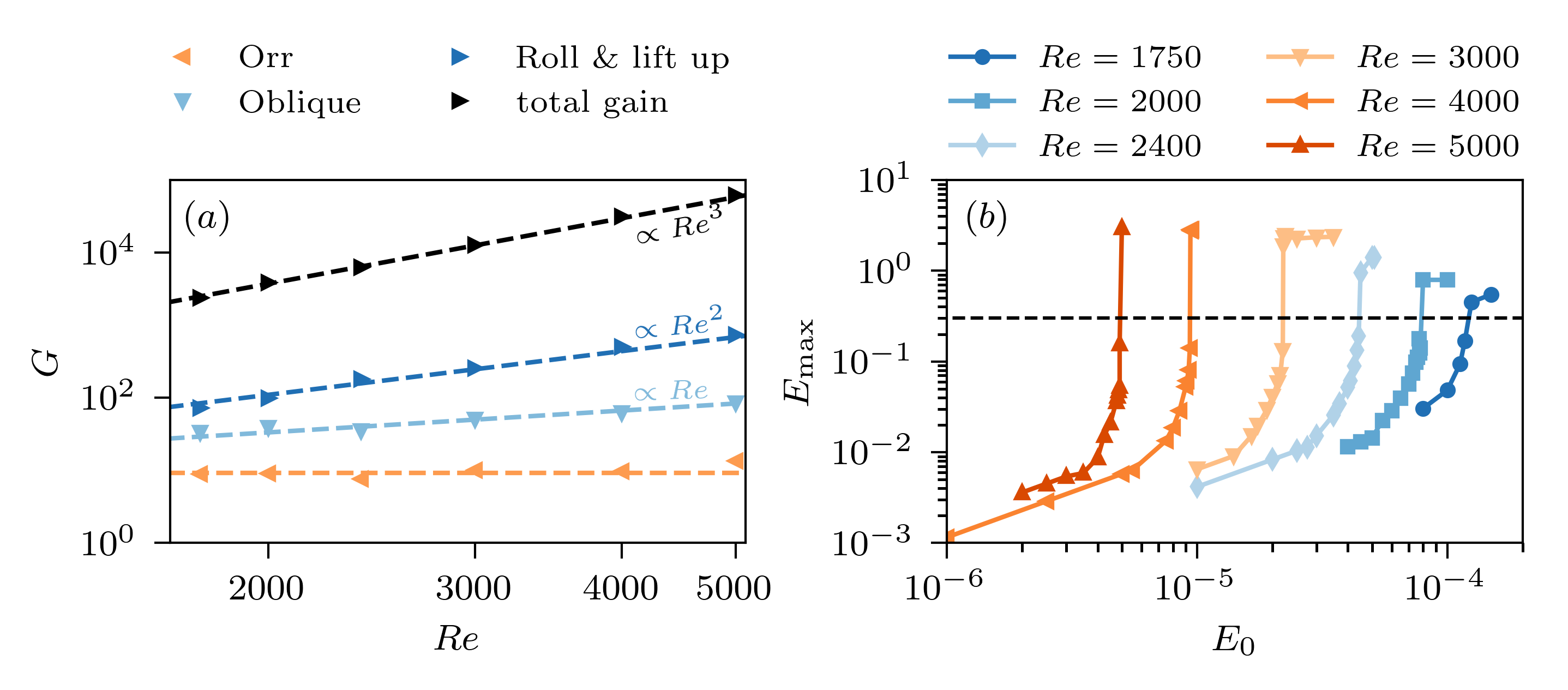}
    \caption{
    $(a)$ The energy gain during the different phase of NLOPs and the overall energy gain as a function of $Re$. 
    $(b)$ The maximum energy achieved by optimal perturbation as a function of their initial energy $E_0$ for various $Re$. The dashed line indicates constant maximum energy $E_\text{max}=0.3$.
    }
    \label{fig: scaling of phases and energy}
\end{figure}

For all $Re$ considered, the initial optimal perturbation retains qualitatively the same structure (figure \ref{fig: ms localisation snapshots}). 
Consistently, the evolution of NLOPs is essentially the same across $Re$. In the following, we analyse the effect of $Re$ on the energy growth of each growth phase.
In figure \ref{fig: scaling of phases and energy}$a$, we show the energy gain of NLOPs after each growth phase together with the overall amplification.
During the initial Orr phase, the energy growth is independent of $Re$ underlining the inviscid nature of the Orr mechanism. Furthermore, the amplification during the Orr phase is independent of $E_0$ so that the Orr mechanism simply constitutes an initial energy boost of $G_\text{Orr} = C_0\approx9.0$.
During the oblique--wave phase, the strongly localised NLOPs rapidly elongate and, although independent of $E_0$, the energy gain increases linearly with the Reynolds number,
\begin{equation}
    G_\text{obl} = \frac{E_\text{obl}}{E_0}\approx C_1 Re,
\end{equation}
where $E_\text{obl}=E(t_\text{obl})$ and $C_1\approx 0.0166$.
Afterwards, the evolution is characterised by strong nonlinear interactions of various oblique modes of relatively large streamwise wave number leading to the growth of rolls and streaks and a further delocalisation of the perturbation. 
In contrast to the oblique--wave phase, the energy growth during the roll and lift--up phase strongly depends on $E_0$ (figure \ref{fig: delta and G(E0)}$b$).
To identify the dependence of the energy gain on $Re$ during these phases, we determine the energy NLOPs require after the oblique phase, $E_\text{obl}$, in order to reach a fixed energy at the end of the lift--up phase, $E_\text{max}$. Here, we compute the required $E_\text{obl}$ for perturbations to achieve $E_\text{max}=0.3$ (see dashed line in figure \ref{fig: scaling of phases and energy}$b$). As we will show later, this corresponds approximately to the edge state energy and optimal perturbations reaching a higher maximum energy will trigger transition. 
At the end of the lift--up phase, we find that
\begin{equation}
    G_\text{(roll \& lift)}=\frac{E_{\max}}{E_\text{obl}}\approx C_2(E_{\max})Re^2,
\end{equation}
where $E_\text{max}=E(t_\text{lift})$ and $C_2(E_\text{max}=0.3)\approx 2.74 \cdot 10^{-5}$.
As the overall energy gain is governed by a sequence of linear inviscid Orr phase, followed by a linear amplification during the oblique--wave phase and a quadratic energy gain during the lift--up phase (see figure \ref{fig: scaling of phases and energy}$a$), the total energy gain of NLOPs increases as
\begin{equation}
    G_\text{max}=\frac{E_\text{max}}{E_0} =C_1C_2(E_\text{max})Re^3.
\end{equation}

\subsection{The transition threshold}
\label{ssec: instability threshold}
For sufficiently large initial energies, $E_0>E_0^\text{crit}$, minimal seeds efficiently trigger transition.
To elucidate the scaling of the transition threshold, $E_0^\text{crit}(Re)$, we determine the lowest initial energy required to trigger turbulence transition, by computing optimal perturbations at various $Re$ and $E_0$. 
\begin{figure}[h]
    \centering
    \includegraphics[width=0.99\linewidth]{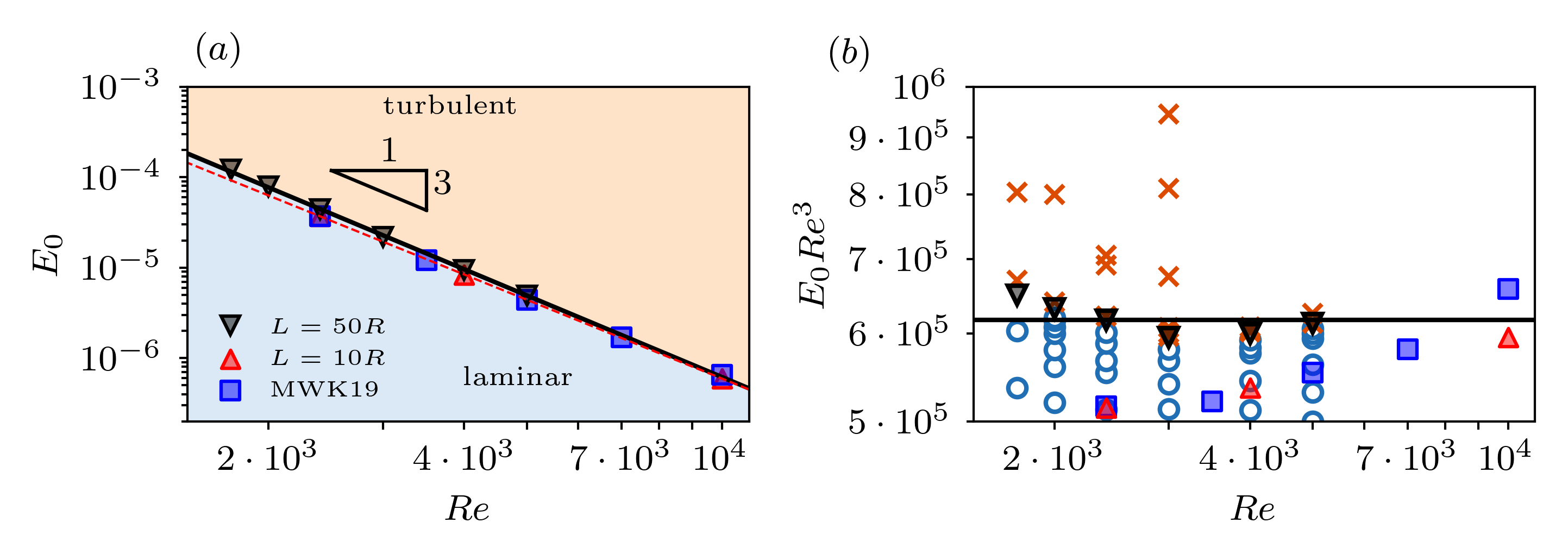}
    \caption{
    $(a)$ The minimal necessary energy $E_0^\text{crit}$ as a function of $Re$ to trigger transition to turbulence. Black and red triangles indicate a linear fit of the minimal seed for $L=50$ and $L=10$, respectively. The solid line corresponds to the transition threshold \eqref{eq: instability threshold}.
    Blue squares indicate the results from \citet{Marensi2019}.
    $(b)$ Same as $(a)$ but compensated by $Re^{3}$ to show the deviation of the results from the scaling at small Reynolds numbers, $Re\lesssim  2000$. Circles indicate NLOPs that do not trigger turbulence transition and crosses correspond to those that trigger turbulent episodes. 
    }
    \label{fig: instability threshold}
\end{figure}
For each optimal perturbation, we compute the maximum growth and verify, if the NLOP lies inside the laminar (blue circles in figure \ref{fig: instability threshold}) or turbulent attractor (red crosses in figure \ref{fig: instability threshold}).
The transition threshold is computed by fitting the critical initial energy, estimated by linear interpolation between the last laminar and first turbulent case (see triangles in figure \ref{fig: instability threshold}) and we find that
\begin{equation}
    \label{eq: instability threshold}
    E_0^\text{crit} \approx C_cRe^{-3},
\end{equation}
where $C_c\approx 616660$.

\subsection{Edge state energy}
From a dynamical systems perspective, initial perturbations of small $E_0$ lie inside the basin of attraction of the laminar state. They can grow transiently, but ultimately decay due to viscous effects.
Perturbations of sufficiently large amplitude, can leverage non--normal growth to reach amplitudes at which nonlinear effects trigger laminar turbulent transition. Such perturbations lie inside the basin of attraction of the turbulent state.
The basin boundary separating the laminar and turbulent attractors is termed the \textit{edge} \citep{Skufca2006}.
The edge constitutes a manifold in state space of perturbations that neither relaminarise nor trigger turbulence, but instead are attracted to the \textit{edge state}, an attractor embedded within the edge \citep{Skufca2006,Schneider2007}.
The exact minimal seed corresponds to the lowest energy state on the \textit{edge} \citep{Kerswell2018}.
Since the energy growth of optimal perturbations increases as $\propto Re^{3}$, equation \eqref{eq: instability threshold} implies that the energy minimal seeds need to reach in order to trigger transition, i.e. the energy of the edge state, is independent of the Reynolds number
\begin{equation}
    E_\text{tr}=
    \underbrace{C_cRe^{-3}}_{E_0^\text{crit}}
    \underbrace{C_1Re}_{G_\text{obl}}
    \underbrace{C_2(E_\text{tr})Re^2}_{G_\text{lift}}
    =  \text{const}.
\end{equation}
To verify the hypothesis, we computed the edge state energy for $ 2000 \leq Re \leq 5000$ via a bisection method whereby the initial energy, $E_0$, is refined until the relative error drops below a tolerance, $(E_0^\text{up}-E_0^\text{low})/E_0^\text{low}<\epsilon$. 
\begin{figure}[h]
    \centering
    \includegraphics[width=0.99\linewidth]{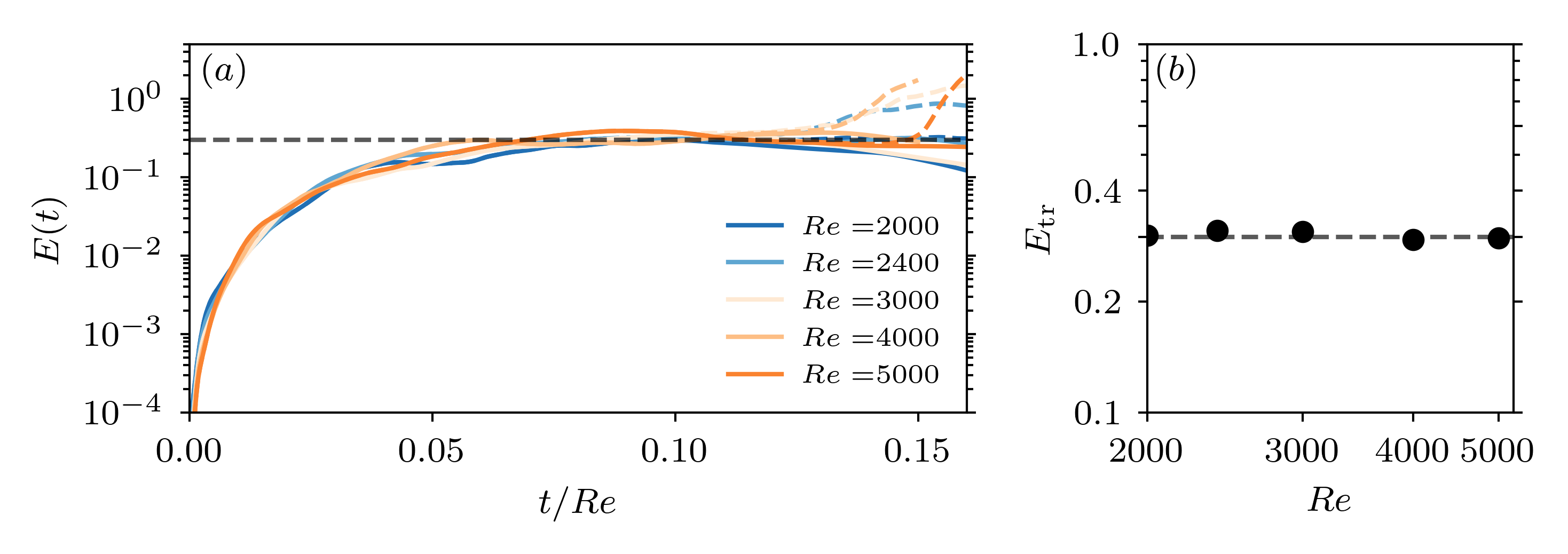}
    \caption{
    $(a)$ The perturbation energy as a function of time of the laminar and turbulent trajectories for different $Re$. The black dashed line indicates the approximated edge state energy, $E_e\approx 0.3$.
    $(b)$ The estimated edge state energy for the different $Re$.
    }
    \label{fig: edge energy}
\end{figure}
To keep the precision of the computed edge state energy at higher $Re$, we adjusted $\epsilon$ accordingly.
We define the edge state energy as the perturbation energy $E(t_e)$, where $t_e$ corresponds to the time at which the relative energy difference of the laminar and turbulent trajectory exceeds $0.5\%$.
For the considered cases, we find the edge state energy is almost constant, $E_e\approx 0.3R^3U_c^2$ (figure \ref{fig: edge energy}$b$), consistent with the scaling of the energy growth and critical initial energy of minimal seeds. Our results agree with \citet{Mellibovsky2009_PRL}, who found that $E_e$ does not change noticeably within $1800\leq Re \leq 2800$ (see their figure 3$a$).

\subsection{The structure of minimal seeds}
\begin{figure}[h]
    \centering
    \includegraphics[width=.99\linewidth]{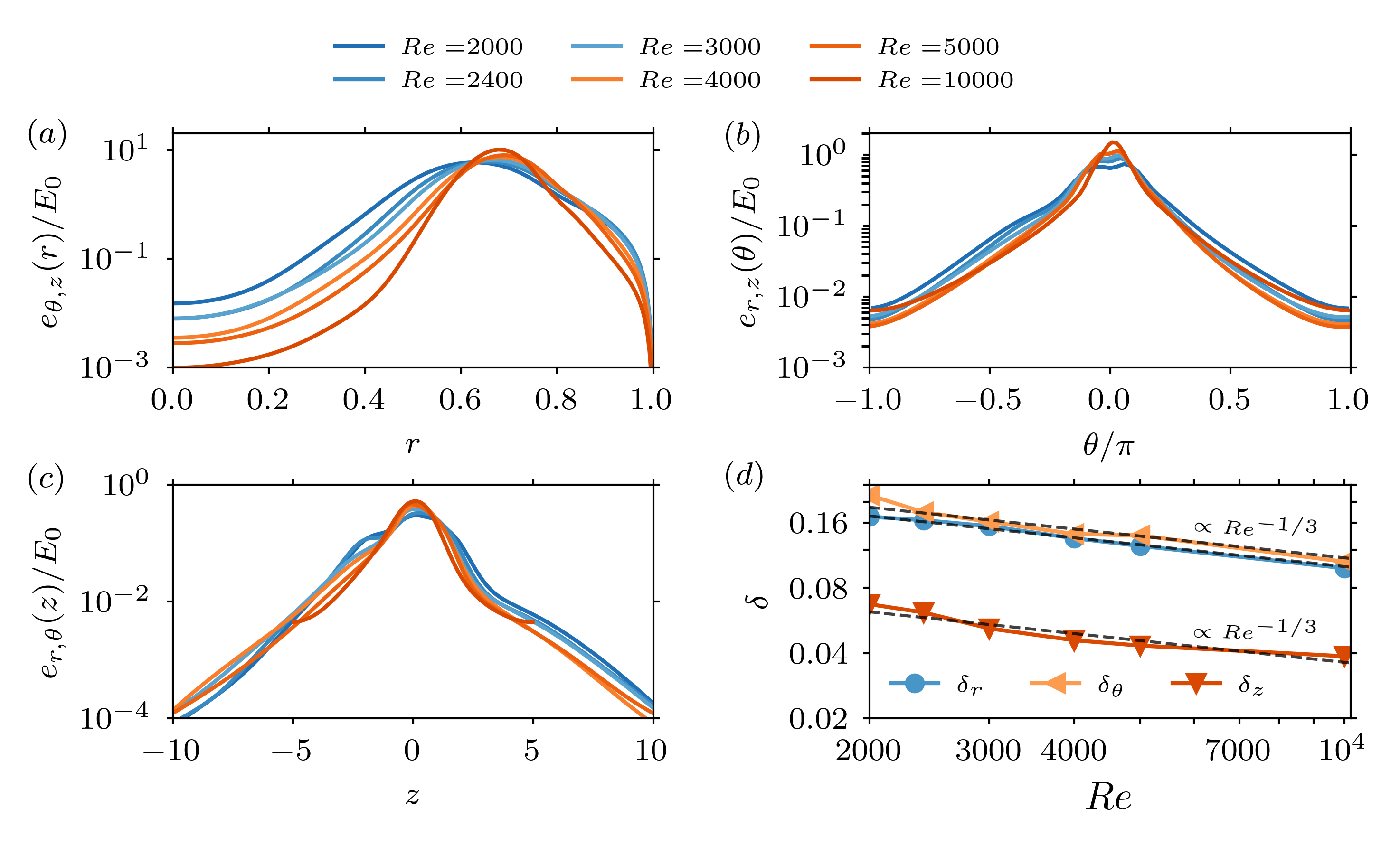}
    \caption{
    $(a)-(c)$ Relative energy density of the NLOP with the highest initial energy that still lies inside the laminar attractor as a function of $r, \ \theta$ and $z$ for different $Re$. In the periodic directions, $\theta,z$, NLOPs are shifted so that their centre of mass of the energy density lies at the centre. For $e_{r,\theta}(z)$ only a $20R$ long section of the $L=50$ pipe is shown.
    $(d)$ The relative thickness of the initial perturbation, $\delta$, in each direction. Dashed lines are $\propto Re^{-1/3}$.
    }
    \label{fig: ms localisation scaling}
\end{figure}

As $Re$ increases, the minimal seed remains structurally similar, but further localises in all three directions (figure \ref{fig: ms localisation scaling}$a-c$).
This allows perturbations to still locally leverage nonlinearities despite the decreasing initial energy.
To quantify this increasing degree of localisation, we compute the energy density as a function of the respective spatial coordinate and the corresponding relative spatial thicknesses
\begin{equation}
    \delta_r=\int\limits_0^{1} \frac{e_{\theta,z}(r)}{\max\limits_r(e_{\theta,z})}rdr,
    \hspace{10mm}
    \delta_\theta=\frac{1}{2\pi}\int\limits_0^{2\pi} \frac{e_{r,z}(\theta)}{\max\limits_\theta(e_{r,z})}d \theta,
\end{equation}
in addition to the axial thickness, equation \eqref{eq: axial thickness}.
As $Re$ increases, the radial, azimuthal and axial thickness decreases as $Re^{-1/3}$ (figure \ref{fig: ms localisation scaling}$d$). 
We note that the perturbation for $Re=10^4$ was computed with $L=10$, which may affect the localisation and result in the axial thickness being larger than expected. We will further discuss the case at $Re=10^4$ in \S \ref{sec: Discussion}.

\section{Discussion}
\label{sec: Discussion}
The scaling exponent of $\gamma=-3.0$ obtained in this study is slightly lower than the exponent of $\gamma={-2.8}$ for $L=10$ previously reported by \citet{Marensi2019}. To understand the origin of this difference, we additionally computed minimal seeds for $L=10$. We obtained $\gamma\approx {-2.89}$ (see red dashed line in figure \ref{fig: instability threshold}$a$), which indicates that the difference in $\gamma$ stems from the sensitivity of $E_0^\text{crit}$ to $L$.
At low $Re=2400$, where NLOPs are less localised, the difference is about $16\%$. Due to the increasing localisation of NLOPs with $Re$, the difference reduces to $10\%$ at $Re=4000$.
The decreasing axial thickness with $Re$, suggests that $L=10$ suffices to compute the instability threshold for $Re\gtrsim 10^4$, thereby significantly reducing the computational cost of each optimisation. Indeed, we computed the minimal seed at $Re\gtrsim 10^4$ for $L=10$ and find that $E_0^\text{crit}$ falls on top of the fit \eqref{eq: instability threshold} for the data with $L=50$ and $Re\leq 5000$.
In addition, $E_0^\text{crit}$ is much less sensitive to the numerical resolution than $G$ as discussed in \S \ref{ssec: numerical resolution}. To reduce the computational cost, we use the same spatial resolution, adapted to $L=10$, and a time step of $\Delta t=0.0025$.
\\
Our results are in agreement with the experimentally obtained scaling with \textit{push--pull disturbances} \citep{Peixinho2007}, which were designed to trigger localised oblique waves. Consequently, they can leverage both the linear growth scaling of the oblique phase and the quadratic lift--up scaling, which rationalises their identified $\gamma=-3.0$ exponent of the instability threshold for this perturbation type. Jet--type perturbations, on the other hand, induce localised but weakly bent streaks that are solely amplified via the lift--up mechanism. Without taking advantage of an oblique--wave phase, the growth is expected to scale quadratically with $Re$ and, given the constant threshold energy, yield an instability threshold that scales only as $\gamma=-2.0$, as computed by \citet{Hof2003}.
We note that the Orr phase of NLOPs is a feature of the optimisation to reduce the required $E_0$, but since it is independent of $(Re,E_0)$, it does not contribute to the scaling of the instability threshold.

\section{Conclusion}
\label{sec: Conclusion}
In this paper, we computed nonlinear optimal perturbations to steady pipe and examined the effect of the numerical parameters on the optimisation results.
For sufficiently large initial energies, $E_0$, optima are fully localised and leverage the Orr and lift--up mechanisms, and strong nonlinear interactions of oblique waves, to experience a large energy gain.
Qualitatively NLOPs are the same as in shorter pipes \citep{Pringle2010,Pringle2012,Pringle2015} and leverage the same mechanisms as NLOPs in other shear flows like boundary--layer \citep{Cherubini2010, Cherubini2011}, plane Couette \citep{Monokrousos2011, Duguet2010, Duguet2013} or plane Poiseuille flow \citep{Farano2016}.
With increasing $Re$, NLOPs remain qualitatively identical but further localise with thickness scaling as $Re^{-1/3}$.
NLOPs initially grow via the Orr mechanism, with an energy gain independent of both $E_0$ and $Re$, followed by a phase of rapid delocalisation and interaction of oblique waves that scales linearly with $Re$.
Subsequently, NLOPs rely on bent streaks and the sustained nonlinear interaction of oblique waves to sustain cross-sectional vortices and to transfer energy into low wavenumber modes that experience a large long term amplification. Effectively, nonlinearities enhance the growth by the lift--up mechanism. The energy gain strongly increases with $E_0$ and scales quadratically with $Re$. Once NLOPs surpass a critical energy value, $E_\text{max}>E_\text{tr}$, large amplitude streaks generated by the lift--up mechanism become unstable and trigger transition in the form of puffs or slugs. 
Regardless of $Re$, the energy of the edge state approached by the minimal seed is constant, $E_\text{tr}\approx 0.3$. As a consequence, the transition threshold of nonlinear optimal perturbations decreases approximately as $E_0^\text{crit} \propto Re^{-3}$.
\\
As the oblique--wave, roll and lift--up phase are generic features of NLOPs also observed in other shear flows, e.g. in plane Couette and plane Poiseuille flow, the same scaling of the instability threshold is expected provided that the domain is sufficiently long for oblique waves to elongate without self--interaction.
\\
Accurate calculations of the energy gain, require sufficiently long pipes, careful optimisation times and a fine numerical resolution. 
In contrast, to resolve the transition threshold, $E_0^\text{crit}$, a much lower numerical resolutions suffices. Similarly, $E_0^\text{crit}$ is less sensitive to the choice of the optimisation time.
The pipe length, on the other hand, has a substantial effect also on the transition. At higher $Re$, NLOPs increasingly localise in the streamwise direction. Comparing the transition threshold for different pipe lengths, suggest that $L=10$ suffices for $Re \gtrsim 10^4$, whereas $L=50$ is needed at $Re\lesssim4000$.
\\
Future work towards predicting transition in practical systems could address transition induced by coloured noise with different correlations in space and time (similar to \citet{Frame2024}).
Comparing flow features of such suboptimal transition routes to NLOPs might reveal statistically dominant mechanisms and could contribute towards control techniques in systems where instantaneous perturbation structures are unknown but whose statistics may be prescribed.

\section*{Acknowledgement}\label{Ack}
\noindent
This work was funded by the Deutsche Forschungsgemeinschaft (DFG) in the framework of the research unit FOR 2688 'Instabilities, Bifurcations and Migration in Pulsatile Flows' under grant number 349558021, which is gratefully acknowledged.
\section*{Deceleration of Interest} 
The authors report no conflict of interest.
\section*{Author ORCIDs} 
\orcidlink{0009-0005-6141-3145} Patrick Keuchel \href{https://orcid.org/0009-0005-6141-3145}{\textcolor{blue}{https://orcid.org/0009-0005-6141-3145}}; \\
\orcidlink{0000-0001-7057-0082} Daniel Mor\'on Montesdeoca \href{https://orcid.org/0000-0001-7057-0082}{\textcolor{blue}{https://orcid.org/0000-0001-7057-0082}}; \\
\orcidlink{0000-0001-5988-1090} Marc Avila \href{https://orcid.org/0000-0001-5988-1090}{\textcolor{blue}{https://orcid.org/0000-0001-5988-1090}}.
\bibliographystyle{jfm}
\bibliography{jfm.bib}
\end{document}